\documentclass[prl,twocolumn,superscriptaddress]{revtex4-1}

\usepackage{epstopdf}
\usepackage{graphicx}

\begin{document}

\title{How fish power swimming: a 3D computational fluid dynamics study}

		\author{
Tingyu Ming}

\affiliation{
		Beijing Computational Science Research Center \\
		Haidian District, Beijing 100193, China }
		
		\author{
Bowen Jin}
\affiliation{
		Beijing Computational Science Research Center \\
		Haidian District, Beijing 100193, China }

		\author{
Jialei Song}
\affiliation{
		Beijing Computational Science Research Center \\
		Haidian District, Beijing 100193, China }
\affiliation{
		Department of Mechanical and Automation Engineering, Chinese University of Hong Kong\\
		Hong Kong SAR, China }

		\author{
Haoxiang Luo}
\affiliation{
		Department of Mechanical Engineering, Vanderbilt University \\
		Nashville, TN 37212, USA}

		\author{
Ruxu Du}
\affiliation{
		Department of Mechanical and Automation Engineering, Chinese University of Hong Kong\\
		Hong Kong SAR, China }

\author{
Yang Ding}\email{dingyang@csrc.ac.cn}
\affiliation{
		Beijing Computational Science Research Center \\
		Haidian District, Beijing 100193, China }

\begin{abstract}
In undulatory swimming of fish, muscles contract sequentially along the body to generate a bending wave that pushes against the water and produces thrust. Here, we use a 3D computational fluid dynamics model coupled to the motion of the fish with prescribed deformation to study the basic mechanical quantities along the fish's body: force, torque, and power. We find that forces on the bodies of both the anguilliform swimmer and the carangiform swimmer are dominated by reactive forces; furthermore, the force on the caudal fin of the carangiform swimmer is dominated by drag-like forces. The torque exhibits a wave pattern that travels faster than the curvature wave in both the anguilliform and carangiform swimmers, but the wave speed is even higher for the carangiform swimmer. The power output for the anguilliform swimmer is concentrated on the anterior half of the body and is significantly negative on the posterior side of the body. In contrast, most of the power is generated by the posterior part of the body before the peduncle for the carangiform swimmer. The results explain the differences in the observed electromyography patterns in fish with different swimming modes and explain the tendon function in carangiform swimmers.

\end{abstract}

\maketitle

\section*{Introduction}
In the undulatory swimming of fish, a backward-traveling wave of body bending is formed to push against the water and generate propulsion. How the thrust is generated has attracted interdisciplinary research in the past few decades, and considerable progress has been made in understanding the propulsion hydrodynamics. Assuming that the force on the body of the swimmer is dominated by the drag and that forces are independently generated by the segments, Taylor developed the resistive force theory (RFT)~\cite{taylor1952analysis}. Perpendicular forces (with a higher drag coefficient than that of parallel forces) on the segments enable propulsion. In the classic elongated body theory (EBT) by Lighthill, the inertia of the surrounding fluid of the body is considered the key effect for propulsion, and the predominant forces on the body are reactive forces, which are required to accelerate/decelerate the water~\cite{lighthill1960note,lighthill1970aquatic}. Considering the inviscid fluid passing by a plate waving with a small amplitude, 2D and 3D waving plate theories were developed~\cite{wu1961swimming,cheng1991analysis}. Experiments using digital particle image velocimetry (DPIV)~\cite{tytell2004hydrodynamics,Nauen2002Hydrodynamics} and bio-inspired robotic studies~\cite{Wen2012Hydrodynamic} suggest that the thrust is generated by the whole body for anguilliform swimmers (e.g. eel) but only the posterior part of the body for carangiform swimmers (e.g. scup and mackerel) . However, inferences of the hydrodynamic stresses and forces on the body from the current data of the velocity field are still unreliable. Comprehensive reviews on the hydrodynamics of fish swimming are given by Sfakiotakis {\it et al.}~\cite{sfakiotakis1999review} and by Lauder \& Tytell~\cite{Lauder2005Hydrodynamics}. Recently, new theories extending and combining reactive and resistive forces have been developed (e.g. \cite{porez2014improved,alben2012dynamics}).

Internally, the swimming of fish is powered by muscles; therefore, how muscles are used is a key question in understanding fish swimming. Electromyography (EMG) measurements on the muscle of fish during swimming provide spatiotemporal patterns of muscle activation in swimming fish of various species (\cite{williams1989locomotion,Wardle1993the,gillis1998neuromuscular,leeuwen1990function,rome1993fish}; for a review, see \cite{altringham1999fish}). During steady swimming, a common pattern emerges: the muscle elements are activated in the manner of a wave traveling posteriorly, but this EMG wave travels faster than the curvature wave. As such, the phase difference between the curvature and EMG varies along the body, known as the ``neuromechanical phase lags''. Nonetheless, details of the muscle activation pattern vary among species. For anguilliform swimmers such as eel, the speed difference is not large and the length of muscle activation on one side of the body is approximately half of the undulation period. For carangiform swimmers such as carp, the propagation speed of EMG onset is much higher than that of the bending wave, whereas that of EMG termination is even higher, resulting in a posteriorly shortening period. The EMG activity, together with  muscle contraction kinetics, the strain and the volume of the active muscle, can determine the absolute muscle power output along the body. Taking this approach, in a paper titled ``How fish power swimming'', Rome {\it et al.}~\cite{rome1993fish} showed that for scup, the power is generated mostly by the posterior part of the body.

To understand the underlying mechanical principles of internal driving and muscle functions, previous studies have used swimming kinematics as input and computed the torque and power required along the body. Using resistive forces and kinematics from leeches, Chen {\it et al.}  found that the torque wave travels faster than the curvature wave, similar to the EMG observation~\cite{chen2011mechanisms}. They also found that the middle part of the body contributes most of the work and that almost no negative work is observed along the body. The torque distribution was calculated by Hess and Videler~\cite{hess1984fast} using elongated body theory and by Cheng {\it et al.} using 3D waving plate theory \cite{cheng1998continuous,cheng1994bending}. The torque pattern obtained by Cheng {\it et al.} also shows a faster traveling wave pattern than the curvature wave, but the phase of the torque lags behind the EMG signal. Since positive and negative torques both occupy half of the period all along the body, shortening of the EMG in carangiform swimmers remains an obscure phenomenon. Another approach to understanding the internal powering in the coupled system is to use neural control signals as input and observe the kinematics emerging from the coupling of internal driving, the body, and the external fluid~\cite{mcmillen2008nonlinear}. Using 2D computational fluid dynamics (CFD) with a prescribed muscle activation, Tytell {\it et al.}~\cite{tytell2010interactions} studied a lamprey-like swimmer and showed that the same muscle forces can generate body bending with different wavelengths, corresponding to different magnitudes of the neuromechanical phase lags, depending on passive body properties such as stiffness.


\begin{figure}[htp]
\centering
\includegraphics[width=0.5\textwidth]{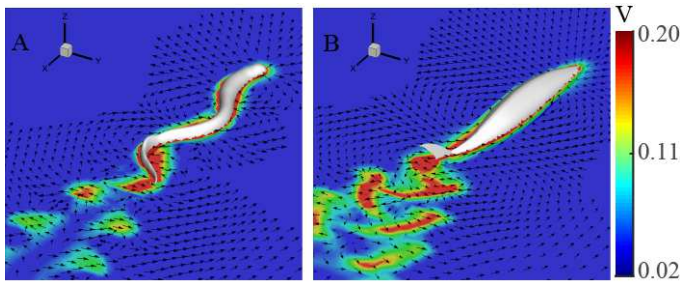}
\caption{Flow fields in the middle coronal ($z=0$) planes of the eel (A) and mackerel (B) models in the lab frame. The arrows represent the velocity direction and the colors represent the magnitude of the velocity. Only flow speeds greater than 0.02 in nondimensionalized units are shown.}
\label{vortex}
\end{figure}

While 3D CFD have been used to study many aspects of fish swimming, previous studies on the internal torque and power are all based on either theoretical models with strong assumptions or 2D CFD models, which cannot capture some 3D effects~\cite{wolfgang1999near} and 3D body shapes. In this study, we use 3D CFD simulations to investigate the force distribution for a typical anguilliform swimmer and a typical carangiform swimmer and to evaluate the theoretical models. We show how the different 3D body shapes and kinematics affect the forces exerted on the body and how such forces lead to different torque patterns and power output patterns. The biological implications of the results are also discussed.

\section*{Results}
\subsection*{Basic Characteristics} Treating the water as an incompressible viscous fluid and the fish as ``rigid'' bodies with prescribed body deformation, we developed numerical 3D models of an eel and a mackerel (see Materials \& Methods for the details). The free swimming speeds ($U$) are 0.29 and 0.25 in nondimensionalized units for the eel and the mackerel, respectively. The corresponding Strouhal numbers are 0.63 and 0.68 . These values are consistent with previous numerical studies at similar Reynolds numbers ($\emph{Re}\approx$4000) (e.g., \cite{borazjani2010role}). For both fishes, double row vortices are shed behind the tail, similar to previous numerical results (see Fig.~\ref{vortex}). The velocity field behind the mackerel clearly shows a backward flow, while a mean flow behind the eel in the fore-aft direction is not easily detected.

\begin{figure}[htp]
\centering
\includegraphics[width=0.5\textwidth]{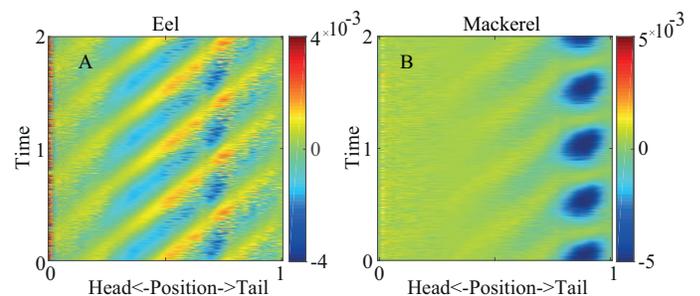}
\caption{Spatiotemporal distribution of the fore-aft force ($F_x$) on the eel (A) and the mackerel (B) for two periods. Negative values indicate thrust as the swimming direction is in the -$x$ direction.}
\label{fx}
\end{figure}

\subsection*{Force} 
\begin{figure*}[htp]
\centering
\includegraphics[width=1\textwidth]{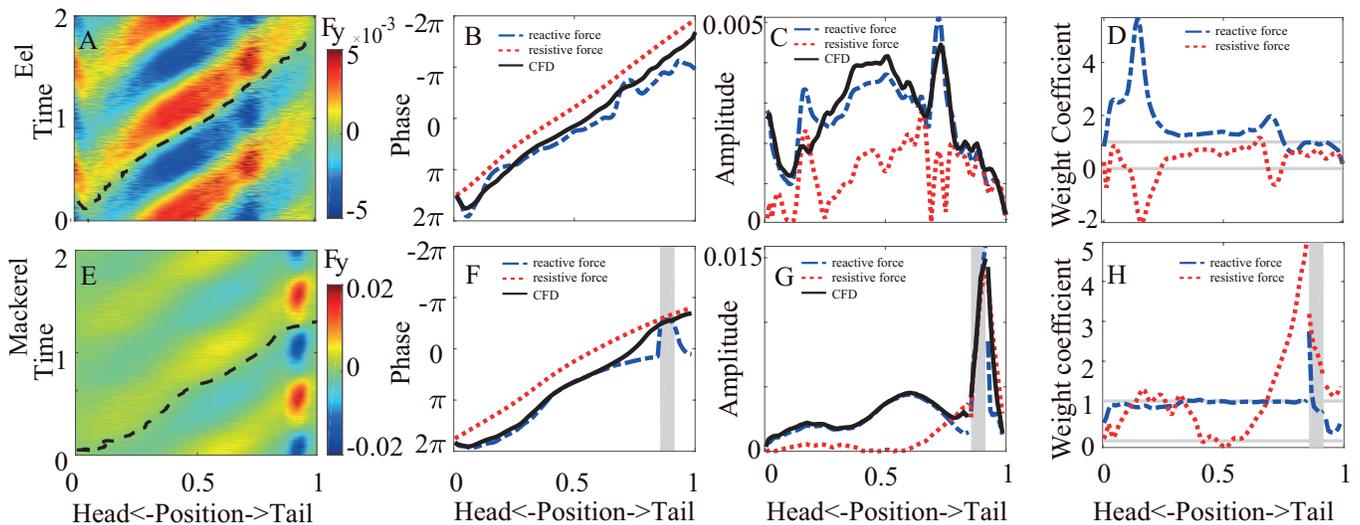}
\caption{Lateral force ($F_y$). Top row: eel; bottom row: mackerel. (A\,\&\,E) Spatiotemporal distribution of the lateral force on the body for two periods from simulation. The dashed black line indicate a zero-crossing (phase) of the force. (B\,\&\,F) Comparison of the phases of the lateral force along the body computed from the reactive force (EBT, blue dashed line), the resistive force (orange dotted line), and CFD (solid black line, as shown in the dashed lines in (A\,\&\,E)). A $2\pi$ term is added or subtracted to ensure continuity. (C\&G) Lateral force computed from the simulation and resistive and reactive forces decomposed from the lateral force from the simulation. The gray area indicates the region where the decomposition is replaced by individual fittings (See text for details).  (D\,\&\,H) Weight coefficients of the resistive and reactive forces. The gray lines indicate 0 and 1 to guide the eye. The symbol schemes are the same across the subfigures except for A \& E.}
\label{force}
\end{figure*}

As expected from the input kinematics and body shapes, the forces are relatively uniformly distributed on the eel but concentrated on the tail of the mackerel (Fig.~\ref{fx}, Fig.~\ref{force}A\,\&\,E and Movie S1\,\&\,S2). The fore-aft and lateral forces both show posteriorly traveling wave patterns that are similar to those of the body bending, except at the head where the surface orientation rapidly changes (Fig.~\ref{fx}). For the eel, the peaks in the force components near 0.7 body length correspond to the bump in the body height at that position. For the mackerel, the separation of thrust and drag is clear: the tail generates most of thrust, and the anterior part of the body generates drag at all times.

To understand the force pattern, we first compare the phase of the lateral force from the simulation with that of the resistive force from RFT (Eq.\ref{RFTF} in M\&M) and the reactive force from EBT (Eq.\ref{EBTF} in M\&M) (Fig.~\ref{force}). Since the resistive force is in phase with the body velocity and the reactive force is in phase with the acceleration of the body acceleration when the body height is constant, the phase of the resistive force lags the reactive force by $\approx\pi/2$ in most positions. Overall, the phase of the observed lateral force on the body is close to the phase of the reactive force except near the snout tips and the tail for the mackerel.  At both the snouts and the mackerel tail before the forking, the phase of the reactive force, resistive force, and measured force all nearly collapse. This is because at these regions, the reactive force arises mainly from the increase in the added mass with the lateral velocity; thus, the reactive force is in phase with the velocity rather than the acceleration of the body. Near the two regions at 0.2 and 0.7 of the eel body, the phase of the reactive force moves toward that of the resistive force due to the increase in the height along the body, but the associated shift in the simulation force is much smaller. A similar but insignificant discrepancy also occurs at the middle of the mackerel. Near the peduncle and posterior to the forking of the tail of the mackerel, the phase of the lateral force transitions from the phase of reactive force to the phase of resistive force.

Furthermore, we decompose the lateral force from the simulation into the resistive force and reactive force and examine their contributions (weights) and magnitudes. Because the phases of the observed force, reactive force, and resistive force nearly collapse before the forking of the tail,  decomposition cannot be performed. As an alternative, we fit the lateral forces from CFD to the two theories separately and draw the coefficients on the same plots in Figure \ref{force}. We find that, in the middle part of the body, where the body is relatively uniform and smooth, the forces on both species are mainly attributed to the reactive forces, and the weight coefficients are close to a constant not far from 1. Near the two regions at 0.16 and 0.7 of the eel body, where the height is increased and the measured phase deviates from the theoretical predictions, the weight coefficient of the reactive force becomes significantly greater than 1, and the resistive weight coefficient becomes negative. The contribution from the resistive force is more substantial for the eel than for the mackerel, since the undulation amplitude on the bulk part of the body of the mackerel is small. Anterior to the peduncle of the mackerel, the force gradually transitions from reactive force to resistive, and the weight coefficient for the resistive force becomes much greater than 1. The weight coefficients of both forces at the leading edge of the mackerel tail are much greater than one, consistent with previous results that the vortices shed from the anterior part of the body can enhance the force and thrust production of the tail~\cite{wolfgang1999near,liu2017computational}. While the two types of force at the tail of the mackerel before the forking have nearly the same phase, the coefficient of the reactive force for individual fitting is closer to 1. Therefore, the large force in this region can be explained by either the EBT or enhanced drag force due to vortex-fin interaction.

\begin{figure}
\centering
\includegraphics[width=0.5\textwidth]{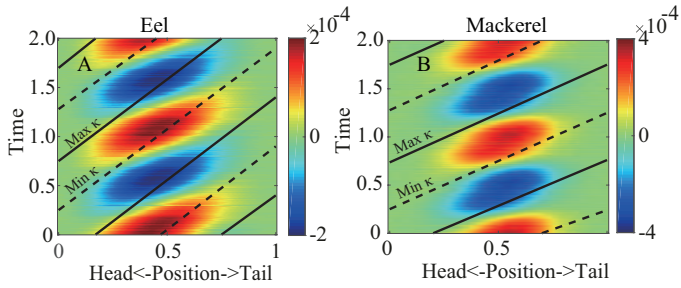}
\caption{Spatiotemporal distribution of the torque on the body in two periods for eel (A) and mackerel (B). The solid and dashed lines indicate the maximum and minimum curvatures, respectively. The same information is illustrated by Movie S3\,\&\,S4.}\label{torque}
\end{figure}

\subsection*{Torque} Considering the aforementioned hydrodynamic forces, we compute the internal torque required to overcome the hydrodynamic forces and body inertia. Body elasticity and other internal resistive forces are initially ignored because quantification of their contributions is still unreliable. The torque in both species exhibits a traveling wave pattern moving posteriorly with higher speeds than the curvature wave speed (Fig.~\ref{torque}). The maximal value of the torque appears at approximately the middle of the body of the eel and slightly posterior to the middle point for the mackerel. The traveling wave speed of the torque is higher in the mackerel than in the eel, exhibiting a nearly standing wave pattern as experimentally observed previously~\cite{wardle1995tuning}.

\subsection*{Power} 
\begin{figure}
\centering
\includegraphics[width=0.5\textwidth]{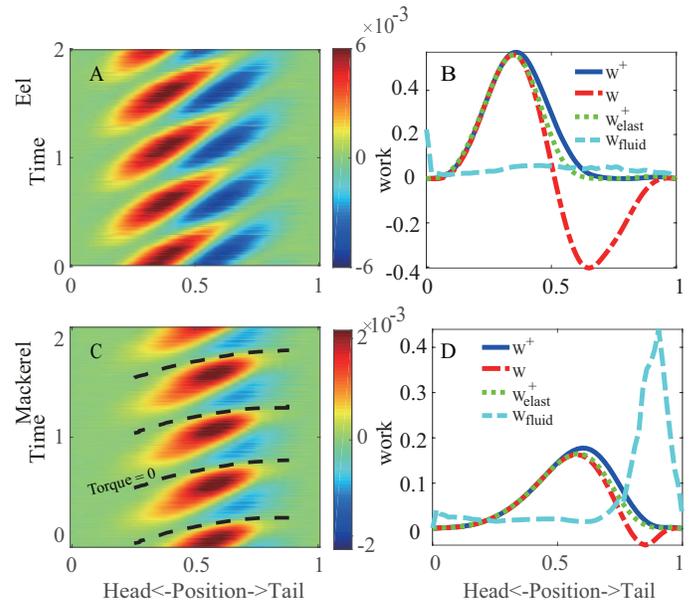}
\caption{Power distribution for the eel (A) and the mackerel (C) and the work done over a cycle by muscles along the body of the eel (B) and the mackerel (D). The dashed line in (C) indicates the zero-crossing of the torque in the mackerel (Fig.~\ref{torque}B). The solid blue lines represent the work by integrating only the positive values ($W^+$) in (B) \& (D), and the dashed-dotted red lines represent the work ($W$) by integrating both positive and negative values. The dotted green lines represent the positive work $W^+$ when body elasticity is considered  (Fig.~\ref{elast}B\,\&\,D). The cyan dashed lines represent the work done to the fluid. }
\label{power}
\end{figure}

As shown in Figure \ref{power}, the power from torque is mostly positive, indicating energy output from the muscle, but negative values are observed on the posterior parts of both the fish. For the eel, the power is nearly all negative for $x>0.6$, similar to the case with a floppy body in the previous 2D study~\cite{tytell2010interactions}, while for the mackerel, the negative power is intermittent on the posterior part. The power calculated by simply integrating the power over a cycle is the minimal power needed, since the dissipation due to the internal resistance is not included; this method implies that the negative power done to the body is fully stored and recovered. The peak of this power is at the anterior part ($\approx$0.4) for the eel and at a more posterior position for the mackerel ($\approx$0.65), slightly posterior to the peak magnitude of the torque. We find that the total work over a period is significantly negative on the posterior half of the eel body and slightly negative near the tail of the mackerel. If we assume that no energy-storing and transmitting elements exist, the work done by the muscles are the integration of only the positive power. We denote this quantity by $W^+$. The differences between the two kinds of work per cycle is greatest for the posterior part of the eel, indicating that power is lost if no spatial energy transfer is performed inside the eel body. The distribution of power done on the fluid from the body is computed from the total force and velocity of a segment (cyan dashed lines in Fig.\ref{power}B\,\& \,D). The power is relatively uniform on the eel but concentrated on the tail of the mackerel.


The mean total power $P_\mathrm{tot}$ averaged over a cycle is $2.0\times 10^{-4}$ (in nondimensionalized units) for the eel and $2.5\times 10^{-4}$ for the mackerel. If only the positive power is used, the power becomes $P_\mathrm{tot}^+=8.5\times 10^{-4}$ and $P_\mathrm{tot}^+=3.3\times 10^{-4}$ , for the eel and the mackerel, respectively.



\section*{Discussion}

\subsection*{Error associated with \emph{Re}}
The low swimming speeds we observed (relative to those of real animals) are likely due to the low \emph{Re} used in our simulations. However, we argue that the results are qualitatively representative for real adult fish. First, a meta-analysis of previously reported fish swimming data indicates that the transition from the viscous regime to the turbulent regime occurs at a \emph{Re} of several thousands~\cite{weerden2014meta}. Second, even the eel model in our study shows the inertia-dominated mode of swimming. Since drag coefficient decreases slowly with increasing \emph{Re} in general, the speed of the simulated swimmer is expected to increase with increasing \emph{Re} and the contribution of the resistive force is expected to decrease for the real adult eel.

\subsection*{Hydrodynamic force model}
Although increasingly accurate CFD methods such as ours are becoming more available, fast calculations of hydrodynamic forces are useful in cases where CFD is too time-consuming, such as on-line control of robotic fish and design optimization~\cite{Christophe2013On}. Our results show that the classic EBT can provide satisfactory predictions on the force distribution on the body and that supplementing EBT with resistive forces is overall a reasonable approach. The current results also reveal the limitations of the force models and suggest strategies for improvement. First, the term associated with the changes in the body height in the EBT is overestimated. One reason might be the full slip condition on the body assumed in the EBT; specifically, the fluid near the body moves at a relative speed $U$ to the body. However, even though the boundary layer is thin, the effective fluid speed near the swimmer could be significantly slower than $U$ because of the the large undulation amplitude, particularly for anguilliform swimmers. Second, the momentum change due to a rapid decrease in the body height, e.g., near the tail of the eel or peduncle of the mackerel, should be reconsidered. In the EBT, the momentum of the decreased part of the added mass is fully transferred to the body. Such effect is responsible for the deviation (advancement) of the phase of the reactive force before the tail (Fig.~\ref{force}F). However, such momentum transfer is unlikely to be perfect since the surface area that interacts with the fluid is disappearing in these regions. Evidently, the phase of the force from the simulation does not exhibit an advance but a lag. Third, the large magnitude and variations of the coefficients of the forces near the peduncle and the tail of the mackerel indicate that the unsteady effects and vortex-body interaction are significant.

\begin{figure}
\centering
\includegraphics[width=0.5\textwidth]{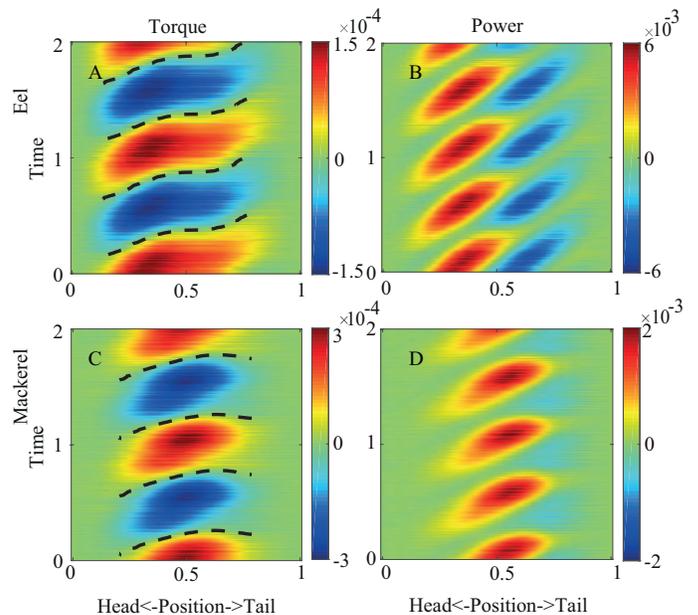}
\caption{Torque (left column) and power (right column) distributions when elasticity of the body is considered for the eel (top row) and the mackerel (bottom row). The dashed lines indicate zero-crossings.}
\label{elast}
\end{figure}

\subsection*{Understanding the torque and power patterns}
The torque pattern can be understood by applying the results obtained in a previous study: The torque pattern in undulatory locomotion is determined mainly by the wavelength and the phase of the lateral force relative to the lateral velocity~\cite{ming2018transition}. The torque wave of the eel has a relatively low wave speed relative to the case of the mackerel due to the short wavelength of undulation. Since the phase of the force for the eel is between those of the resistive force and reactive force, the internal torque and power patterns are between those patterns associated with pure reactive force and pure resistive force (SI Appendix). For the mackerel, the long wavelength of the curvature wave and the concentrated force on the tail result in nearly synchronized torques on the body. Because the force from the tail to the fluid is nearly in phase with the velocity, the rate of change of curvature ($\dot{\kappa}$) and the torque are also nearly in phase. Therefore, the internal power is nearly all positive.

\subsection*{Body elasticity \& Temporal energy transfer}
The elasticity of the body can influence the torque and power patterns, but accurate \emph{in vivo} measurement of the body elasticity distribution is not available. Therefore, we discuss here the trend of the influences of the body elasticity when the elasticity is small relative to the torque from hydrodynamics and body inertia (Fig.~\ref{elast}). We still use the torque from simulation with prescribed kinematics and assume that the magnitude of the torque from the body elasticity is 20\% of that torque at individual positions along the body, namely $T_e=0.2 \langle T \rangle \kappa(s,t)/ \langle \kappa \rangle $, where ``$\langle \rangle$'' means standard deviation over time. We find that the effect of elasticity on the torque is different along the body, separated by a position ($\approx 0.5$ for the mackerel and $\approx 0.2$ for the eel) where $T$ and $\dot{\kappa}$ are in phase and where power is all positive. Anterior to that point, the torque magnitudes increase and wave speeds decrease; posterior to that point, the torque magnitude decreases, the speed of the torque wave increases. For the eel, the effects of the speed increases ends when the maximal curvature coincide with the minimal torque without elasticity. For the mackerel, the torque wave can even reverse when the phase shift effect of the elasticity is strong. The reversal of wave resembles the reversal of the wave of the offset of the EMG observed on some carangiform swimmers~\cite{wardle1995tuning}. As a result of changes in the torque, the area of the negative power region in the posterior part of the body decreases, and $W^+$ decreases. This observation is consistent with the findings of previous studies that suitable elasticity can save and restore energy to improve efficiency (e.g., \cite{alben2012dynamics}).

\subsection*{Tendon \& Spatial energy transfer}
While local elasticity can transfer energy temporally, spatial transmission of the energy can only be enabled by other structures. In animals, coupled joint articulation by tendons over two or more joints is common and is an effective structure to save and transfer energy~\cite{junius2017biarticular}. For carangiform swimmers, long tendons exist that span over many vertebra~\cite{shadwick2005structure}. We hypothesize that these long tendons are used to transfer energy from the posterior part to the middle part of the body when the negative power appears on the posterior part. This hypothesis can explain the observed shortening of the muscle activation period posteriorly among the carangiform swimmers, including some detailed features: the posteriorly increasing negative power period from the middle of the body matches the decreasing EMG period. The start of the positive power is aligned with the sign change of $\dot{\kappa}$ (the lines in Fig.~\ref{torque}B), resulting in the low speed the same as the curvature wave. The end of the positive power is aligned with the sign change in the torque (the dashed lines in Fig.~\ref{power}B), resulting in the high speed the same as the torque wave. Such differences in wave speeds qualitatively match the onset and offset wave speeds of the EMG. Note that this hypothesis does not contradict the common view that force and energy are transmitted to the tail to interact with the fluid. Actually, the torque is still required when the power is negative on the posterior region and can be provided by the muscle in a more anterior position connected by the tendon. This hypothesis is also consistent with the observation that the EMG period is nearly half the undulation period on the whole body of anguilliform swimmers, which do not possess long tendons~\cite{shadwick2005structure}.

In summary, our 3D numerical model shows that the EBT is generally valid for predicting the force distribution for both anguilliform and carangiform swimmers, although some corrections are needed and the details of the hydrodynamics and force on the peduncle and crescent-shaped tail need further elucidation.  For the carangiform swimmer, due to the resistive (or alike) forces in the tail region and short wavelength of the bending wave, the torque wave has a high wave speed, and the power is mostly positive. In contrast, the negative power is significant on the posterior port of the body for anguilliform swimmer. Our results, combined with biological observations, may explain the different patterns of muscle activation and the presence of long tendons in carangiform swimmers.

\section{Methods}
\subsection*{Body shape \& kinematics}

The carangiform body is modeled after the actual anatomy of a mackerel, whereas the anguilliform body is created from a lamprey computed tomography (CT) scan (see \cite{borazjani2010role} for details). Except for the caudal fin, other fins are neglected for the swimmers. The lengths of the fish bodies ($L$) are used as the unit length in their respective simulations. The bodies are meshed with triangular elements, and some sharp and small structures from the scan are removed to avoid instability of CFD computation. After obtaining the surface data of the two fishes, we reshaped the fishes and re-meshed the surface grid so that our code could accommodate the boundary between the fish and fluid. The sharp and thin tail of the mackerel is modeled as a zero-thickness membranous structure. The number of surface mesh points is 3962 for the eel and 2127 for the mackerel (including 1962 for the mackerel's body and 165 for the tail). See SI Appendix for mesh details. The body mass ($M$) was computed by assuming a uniform distribution of density that is equal to the fluid density and is 1 in nondimensional units. $M=0.0019$ for the eel and $M=0.0101$ for the mackerel.

The kinematics for undulatory locomotion is generally in the form of a posteriorly traveling wave with the largest wave amplitude at the tail. To describe the deformation of the fish bodies, the centerline curvatures $\kappa$ are prescribed in the form of $\kappa(s,t)=A(s)\sin(k s-\omega_u t)$, where $s$ is the arc length measured along the fish axis from the tip of the fish head, $A(s)$ is the amplitude envelope of curvature as a function of $s$, $k$ is the wave number of the body undulations that corresponds to a wavelength $\lambda$, and $\omega_u$ is the angular frequency. We use the undulation period as the unit of time, so $\omega_u=2\pi$. The amplitude envelope $a(s)$ for the anguilliform kinematics has the form of $a(s)=a_\mathrm{max} e^{s-1}$, where $a_\mathrm{max}$ is the tail-beat amplitude. For carangiform kinematics, the amplitude envelope has the form of $a(s)= a_0+a_1s+a_2s^2$. The parameters for $A(s)$ were adjusted to fit the envelope of the movement of real fish observed in experiments~\cite{tytell2004hydrodynamics,hess1984fast}. The parameters used were $a_\mathrm{max}=11.41$, $k=2\pi/0.59$, and $\omega=2\pi$ for the anguilliform swimmer and $a_0=1$, $a_1=-3.2$, $a_2=5.6$, and $k=2\pi/1.0$ for the carangiform swimmer. To avoid generating spurious forces and torques in the interaction between the fish bodies and fluid, we added rotation and translation in the body frame of the swimmers to ensure that the movement of the bodies without external forces satisfies the two conservation laws: linear momentum and angular momentum conservation (see SI Appendix for details).

\subsection*{CFD \& fluid-structure interaction}

The in-house immersed boundary method code used is capable of simulating 3D incompressible, unsteady, and viscous flow in a domain with embedded complex objects including zero-thickness membranes and general 3D bodies~\cite{song2014three,luo2012numerical}. The flow is computed on a nonuniform Cartesian grid in the $x'y'z'$ coordinates. The fluid domain has a size of $8.5\times5\times5$, and a total of $620\times400\times400\approx 99$ million points are used.  The grid is locally refined near the body with the finest spacing at $0.005\times0.005\times0.005$.  The fish models are placed in the center of the computation domain and the body centerlines are in the $z'=0$ plane. A homogeneous Neumann boundary condition is used for the pressure at all boundaries. The flow speed of the inlet flow and outlet flow at the front and back boundaries are set as the swimming speed of trial runs so that the model swimmers move only minimally in the computational domain. The zero-gradient boundary condition is used at all other boundaries.

We considered the fluid-structure interaction in the plane of undulation. Because the body of the swimmers is deforming, the governing equation for the angular degree of freedom is $\mathrm{d}(I\omega )/\mathrm{d}t=T_\mathrm{tot}$, where $I$ is the moment of inertia, $\omega$ is the angular speed, and $T_\mathrm{tot}$ is the total hydrodynamic torque. Since the deformation is prescribed, $I$ and $\dot{I}$ are known. Therefore, $\omega$ can be obtained by numerically integrating $\dot{\omega}=(T_\mathrm{tot}-\dot{I}\omega)/I$ while integrating other equations for the translational body movement and the fluid movement. The time interval for integration is $5\times 10^{-4}$.

Since the swimming direction is not perfectly aligned with the $x'$-axis of the computation grid, a new coordinate system is used so that the swimming direction is aligned with -$x$, $y$ is the lateral direction, and the $z$-axis is vertical. The vertical motion is neglected, but the force magnitude in the $z$ direction is only $9.3\times 10^{-6}$ for the eel and $9.3\times 10^{-5}$ for the mackerel, which are less than 3\% of the the force magnitude in the forward direction. The Reynolds number is defined as $Re=UL/\nu_k$, where $\nu_k=1/15000$ is the kinematic viscosity.

\subsection*{Force, torque, and power in simulation}

The force per unit length of fish from the simulation can be calculated as follows: Take an arc length $\Delta s$ along the body centerline, and integrate all forces from every grid point in $\Delta s$; then, divide the total force by the arc length $\Delta s$. The torque required to overcome the hydrodynamic forces and inertia of the body can be computed by integrating the contributions $\mathbf{r}\times (\mathbf{F}-m_b\mathbf{a})\cdot \mathbf{e}_z$ from either side of the body from the point of interest, where $m_b$ is the body mass per unit length and $\mathbf{a}=\dot{\mathbf{v}}$ is the acceleration of the body segment. To minimize the numerical error, we use a weighted average of the torques computed from both sides. The internal power by torque and the power done to the fluid per unit length are computed as $P_T\left( s,t \right)=T\dot{\kappa}$ and $P_F\left ( s,t \right )=-\mathbf{F}\cdot \mathbf{v}$, respectively. The difference between the total power computed by integrating the internal power or external power along the body is within numerical error ($ <5\% $).



\subsection*{Resistive and reactive forces}

The instantaneous resistive force per unit length is computed as 
\begin{small}
\begin{equation}\label{RFTF}
F_{s}\left(s,t\right)=-\frac{1}{2}C_{d}\rho H\left(s\right)\nu^2\left(s,t\right),
\end{equation}
\end{small}
where $\rho$ is the fluid density, $C_{d}$ is the drag coefficient, $H\left(s\right)$ is the body height, and $\nu\left(s,t\right)=\frac{\partial h}{\partial t}$ is the lateral velocity of the midline.  The lateral displacement of the body midline $h$ is extracted from the simulations. In the bifurcated region of the mackerel tail, $H$ is computed as the sum of the heights of both the top and bottom parts of the tail. Approximately considering the cross-sectional shape of the fish bodies, $C_{d}$ is set as 0.5 for the bodies of the eel and mackerel, and 1.0 for the tail of the mackerel. 

In the EBT \cite{lighthill1970aquatic}, the instantaneous reactive force per unit length of fish in the lateral direction is given by 
\begin{small}
\begin{equation}\label{EBTF}
F_a\left(s,t \right )=-\left ( \frac{\partial }{\partial t}+ U\frac{\partial }{\partial x} \right )\left [ V\left ( s,t \right ) m\left ( s \right )\right ],
\end{equation}
\end{small}
where $U$ is the swimming speed, $V\left(s,t\right)= \frac{\partial h}{\partial t} + U\frac{\partial h}{\partial x}$, $m\left(s\right)=\rho\frac{\pi}{4}H^2$ is the added mass per unit length.


The contributions of the reactive force and resistive force are obtained by the best fitting of $F_y(s,t)=C_a(s) F_a(s,t)+C_s(s) F_s(s,t)$, where $C_a$ and $C_s$ are the weight coefficients as a function of position.

\section{Acknowledgment}
This work was supported by the National Science Foundation of China grant No. 11672029 and NSAF-NSFC grant No. U1530401 (both to T.Y.M., B.W.J., J.L.S. \& Y.D.). We thank Prof. Fotis Sotiropoulos and Prof. Iman Borazjani for sharing the shape data of the fish.

\bibliography{torquebib}

\bibliographystyle{apsrev4-1}

\end{document}